\pdfoutput=1

\documentclass[final,5p,twocolumn]{elsarticle}

\usepackage{amssymb}
\usepackage{hyperref}

\biboptions{sort&compress}

\begin{document}

\begin{frontmatter}

\title{Vector - axial vector lattice cross section and valence parton distribution in the pion from a chiral quark model}

\author[ifj,ujk]{Wojciech Broniowski}
\ead{Wojciech.Broniowski@ifj.edu.pl}

\author[ugr]{Enrique Ruiz Arriola}
\ead{earriola@ugr.es}

\address[ifj]{The H. Niewodnicza\'nski Institute of Nuclear Physics, Polish Academy of Sciences, 31-342 Krak\'ow, Poland}
\address[ujk]{Institute of Physics, Jan Kochanowski University, 25-406 Kielce, Poland}
\address[ugr]{Departamento de F\'isica At\'omica, Molecular y Nuclear and Instituto Carlos I de
  Fisica Te\'orica y Computacional, Universidad de Granada, E-18071
  Granada, Spain}

\date{ver. 2, 10 June 2020}

\begin{abstract}
Within a chiral quark model, we evaluate the good cross section in the
Euclidean space in the vector - axial vector channel, proposed recently by Ma and Qiu as means to extract
the so far elusive parton distribution functions of the pion from
lattice QCD. Our results are remarkably simple at the quark model scale and
agree well, after the necessary QCD evolution, with the most recent lattice
calculations at the scale $\mu \sim 2$~GeV for various values of the lattice pion mass.  Comparisons are made as functions of the Ioffe time 
variable. We also comment on the information on the lowest moments in the momentum fraction $x$, generically extractable from such analyses,
as well as on the inaccessibility of the $x \to 1$ limit from the present data.
\end{abstract}


\end{frontmatter}

\section{Introduction}

Hadron structure is most directly visualized in deep inelastic
scattering experiments with a typical momentum transfer $Q \gg
1$~GeV, where the underlying partonic properties or, more
specifically, the quark and gluon composition is unveiled at a given
resolution wavelength $\sim{}1/Q$. Due to this fundamental nature,
numerous attempts have been made over the last 20 years in order to
determine, {\it ab initio} from QCD, the parton distribution functions
(PDFs) and their functional dependence on the momentum fraction
carried by the quarks or gluons, $x$. What is less known is that this
problem was actually faced squarely in the physical Minkowski space in
the so-called transverse Hamiltonian lattice approach for the
pion~\cite{Burkardt:2001mf,Burkardt:2001jg,Dalley:2002nj}. For the
more popular Euclidean space lattices several computational schemes
have been implemented.  They range from the early few moments
determinations of
PDFs~\cite{Martinelli:1987zd,Morelli:1991gb,Best:1997qp,Detmold:2003tm},
to the more recent
quasi-PDFs~\cite{Ji:2013dva,Chen:2016utp,Alexandrou:2015rja,Alexandrou:2017dzj} and
pseudo-PDFs~\cite{Radyushkin:2016hsy,Radyushkin:2017cyf,Orginos:2017kos,Monahan:2017oof,Karpie:2018zaz,Joo:2019bzr,Joo:2019jct,Karpie:2019eiq,Joo:2020spy,Bhat:2020ktg},
the lattice cross sections approach~\cite{Ma:2014jla,Ma:2017pxb}, the
LaMET method~\cite{Ji:2014gla,Chen:2018fwa}, or the Compton
Feynman-Hellman approach~\cite{Chambers:2017dov}.  The necessary
formulation in the Euclidean space, which makes the path integral well
defined, hampers direct extractions based on suitable extrapolations
to the physically accessible Minkowski PDFs. For comprehensive
overviews we refer the reader to the white paper~\cite{Lin:2017snn}
and the review~\cite{Cichy:2018mum}. A recent analysis of quasi distributions in the QCD instanton vacuum model
was presented in~\cite{Kock:2020frx}.

Whereas these different setups aim to be sufficiently accurate to
reliably extract the PDFs at a given scale $\mu$ and compare them to
current phenomenological parameterizations extracted from various
experiments at fixed $Q^2$ values, they in fact propose to calculate
different mathematical objects in QCD on the lattice at a given
spacing $a$. These objects are, however, interesting on their own,
since theoretical hadronic models can be directly tested against them after the
relevant probing scales are consistently tuned, $\mu \sim Q \sim 1/a$,
to judiciously represent a similar physical situation.

This is precisely the aim of this paper, where we compare the lattice
cross section~\cite{Ma:2014jla,Ma:2017pxb} for the pion in the vector
- axial vector channel, $\sigma_{VA}$, obtained
in~\cite{Sufian:2019bol,Sufian:2020vzb}, to the results of the
Nambu--Jona-Lasinio (NJL) model followed with the leading-order (LO)
DGLAP~\cite{Gribov:1972ri,Dokshitzer:1977sg,Altarelli:1977zs}
evolution.  In general, the lattice cross sections, following the early proposal of~\cite{Braun:2007wv}, are a broad class
of objects suited for lattice studies, with the following features
described by Ma and Qiu~\cite{Ma:2014jla}: they are calculable in the
Euclidean lattice QCD, have a well-defined continuum limit, and share
the same and factorizable logarithmic collinear divergences as PDFs.
Our comparison, made for $\sigma_{VA}$ at LO as a function of the
Ioffe-time, shows a comfortable agreement within the error bars in the whole domain of the
lattice data. More generally, we argue that the lattice data for
$\sigma_{VA}$ can be viewed as a determination of the lowest even
valence quark moments $\langle x^n\rangle$, with satisfactory accuracy
up to $n=4$, with no strong sensitivity on the behavior in $x \to 1$
region.

Our strategy follows the earlier works on the pion's
PDF~\cite{Davidson:1994uv,Davidson:2001cc,Weigel:1999pc}, distribution
amplitude~\cite{RuizArriola:2002bp}, generalized distribution
functions~\cite{Broniowski:2007si}, quasi-distribution
amplitude~\cite{Broniowski:2017wbr}, quasi- or
pseudo-PDFs~\cite{Broniowski:2017gfp}, as well as the double
distribution functions~\cite{Broniowski:2019rmu}.  Notably, the pion,
which is the pseudo-Goldstone boson of the dynamically broken chiral
symmetry in QCD, has many of its properties constrained by low energy
theorems. However, it is most challenging from the point of view of
lattice QCD and achieving the low physical value of the pion mass has
always been a tough numerical issue requiring sufficiently large
lattice volumes, such that $e^{-m_\pi V^\frac13} \ll 1$. Also, the
experimental extractions of its partonic distributions are not direct
and require detailed analyses based on QCD factorization. For these
reasons, model studies of sophisticated features of the pion, such as
the one presented here, are useful in illustrating theoretical ideas
and prove helpful to understand the experimental data or the lattice
simulations.

\section{Basic definitions and methodology}

We first very briefly review the relevant definitions and establish the notation.
The starting point is  the so-called {\em lattice cross-section} introduced in~\cite{Sufian:2019bol}
\begin{eqnarray}
\sigma^{\mu \nu}_{ab} ( \xi, p) \equiv \xi^4 \langle \pi (p) | T \left\{ J_a^\mu (\xi) J_b^\nu (0)\right\} | \pi (p)  \rangle, \label{eq:VA}
\end{eqnarray}
where $a,b$ indicate vector ($V$) or axial vector ($A$) currents,
\begin{eqnarray}
J_V^\mu (x) &=& \bar q(x) \gamma^\mu q(x), \nonumber \\
J_A^\nu (x) &=& \bar q(x) \gamma^\mu \gamma^5 q(x). \label{eq:currents}
\end{eqnarray}
The (Euclidean) coordinate $\xi$ separates the two current insertions, whereas $p$ denotes the momentum of the pion. The isospin indices are suppressed for brevity.
One should note that actually Eq.~(\ref{eq:VA}) describes a two-current correlator in the pion state, thus is a genuine 4-point function. 

The Lorentz decomposition of the combination $\sigma^{\mu \nu}_{VA} ( \xi, p)+\sigma^{\mu \nu}_{AV} ( \xi, p)$ is antisymmetric in 
$\mu \nu$, with invariant structures multiplying the tensors $\epsilon^{\mu \nu \alpha \beta}\xi_\alpha p_\beta$ and $\xi^\mu p^\nu - \xi^\nu p^\mu$~\cite{Sufian:2019bol}.
At LO considered here only the former matters, with the coefficient denoted as
$\sigma_{VA} ( \omega, \xi^2 , p^2 )$, where $\omega= p \cdot \xi$ is
the Ioffe time~\cite{Ioffe:1969kf,Braun:1994jq}. It has been shown in Ref.~\cite{Ma:2017pxb} that the following factorization relation holds for $\xi \Lambda_{\rm QCD} \ll 1$:
\begin{eqnarray}
&& \hspace{-1.2cm} \sigma_{VA} ( \omega, \xi^2, p^2) =\int_0^1 \frac{dx}{x} F( x \omega , \xi^2 , x^2 p^2 ;\mu ) \, q_{\rm val} (x ; \mu) \nonumber \\
&& + {\cal O} ( \xi^2 \Lambda_{\rm QCD}^2 ), \label{eq:fact}
\end{eqnarray}
where $\mu$ is the factorization scale, the valence  (non-singlet) PDF of the pion is
\begin{eqnarray}
q_{\rm val} (x;\mu)  = q(x;\mu)- \bar q (x;\mu)
\end{eqnarray}
and $F$ is a perturbative kernel which at LO
and for $\xi=0$ is equal to $x \cos(\omega x)$, yielding Eq.~(35) from~\cite{Sufian:2019bol}:
\begin{eqnarray}
\sigma_{VA} (\omega ) = \int_0^1 \frac1{\pi^2} \cos ( \omega x) \, q_{\rm val} (x ; \mu). \label{eq:fLO}
\end{eqnarray}
We recognize here the real part of the Ioffe-time distribution (ITD) at $\xi^2=0$. ITD is also a focal point of the 
pseudo-PDF studies~\cite{Radyushkin:2016hsy,Radyushkin:2017cyf,Orginos:2017kos,Karpie:2017bzm,Monahan:2017oof,Joo:2019bzr,Joo:2020spy,Bhat:2020ktg},
hence one finds a link between the two methods.
Note that $q_{\rm val} (x;\mu)$ is scale dependent, thus becomes a
function of the renormalization scale $\mu$.

The methodology of~\cite{Ma:2017pxb,Sufian:2019bol,Sufian:2020vzb} is aimed at effectively inverting Eq.~(\ref{eq:fLO}) 
or its NLO version to obtain $q_{\rm val} (x ; \mu)$ 
from the lattice data for $\sigma_{VA} (\omega )|_{\mu}$. In contrast, we proceed with Eq.~(\ref{eq:fLO}) directly, using model 
PDF of the pion in the integrand and confronting the obtained result to the lattice data for for $\sigma_{VA} (\omega )|_{\mu}$.

It is useful to introduce the standard Mellin moments of the valence PDF,
\begin{eqnarray}
\langle x^n \rangle_\mu  = \int_0^1 dx \, x^n q_{\rm val} (x ; \mu) .
\end{eqnarray}
At LO, the dependence on the scale $\mu$ is deduced from 
the solution of the DGLAP equations, which  becomes
very simple for the moments,
\begin{eqnarray}
\langle x^n \rangle_\mu= r^{\gamma_n^{(0)} / 2 \beta_0} \langle x^n \rangle_{\mu_0},  \label{eq:momevo}
\end{eqnarray}
with the {\em evolution ratio} defined as
\begin{eqnarray}
r = \frac{\alpha (\mu)}{\alpha(\mu_0)}.
\label{eq:evol-r}
\end{eqnarray}
Here $\alpha (\mu )=  \frac{4 \pi}{\beta_0} /{\ln \left( \frac{\mu^2}{\Lambda_{\rm QCD}^2}\right)}$ is the LO running coupling constant,
$\Lambda_{\rm QCD}=226$~MeV, $\beta_0= \frac{11}{3} N_f - \frac{2}{3} N_c$ ($N_f=3$), and $\gamma_n^{(0)} $ are the
LO non-singlet anomalous dimensions,
\begin{eqnarray}
\gamma_n^{(0)} = -2 C_F \left(-4 H_{n+1}+\frac{2}{(n+1) (n+2)}+3\right)
\end{eqnarray}
with $C_F= 4/3$, and $H_{n+1}= \sum_{k=1}^n 1/k $ denoting the harmonic
sum. When  needed, the PDF can then be reconstructed from the moments by means of an inverse Mellin
transform after analytic continuation to the complex $n$ plane
(see, e.g., Ref.~\cite{RuizArriola:1998er} for details).

If we proceed by a power series expansion for small $\omega$, we get 
\begin{eqnarray}
\pi^2 \sigma_{AV}(\omega) |_\mu &=& \sum_{n=0}^\infty \frac{(-\omega^2)^n}{(2 n)!} \langle x^{2n} \rangle \nonumber \\  &=&
\sum_{n=0}^\infty \frac{(-\omega^2)^n}{(2 n)!} \langle x^{2n} \rangle_{\mu_0}
r^{\gamma_n^{(0)} / 2 \beta_0}.
\end{eqnarray}
This simple formula allows one to determine $\sigma_{VA}$ at a scale $\mu$,
{provided} we know it at a {\it reference} scale $\mu_0$.
A typically used assignment of the lattice scale $\mu \sim 1/a$ yields $\mu=2$~GeV for the lattice spacing $a=0.1$~fm.

The data for $\sigma_{AV}$ determined on the lattice and displayed
later in Figs.~\ref{fig:sVA} or \ref{fig:comb} exhibit a sizable dependence on the Ioffe time
$\omega$, but simultaneously a very weak dependence on the $\xi^2$
variable. Actually, the coefficient of the $\xi^2$ term extracted from
a fit where finite volume and pion mass effects are also discerned is
compatible with zero. This result is to be expected, as these terms
correspond to higher twist contributions which within the operator product expansion
are connected to the gluon and quark condensates, with non-vanishing
leading contributions starting at ${\cal O} ( \xi^4)$.

\section{Chiral quark model results}

\begin{figure}[tb]
\begin{center}
\includegraphics[angle=0,width=0.38 \textwidth]{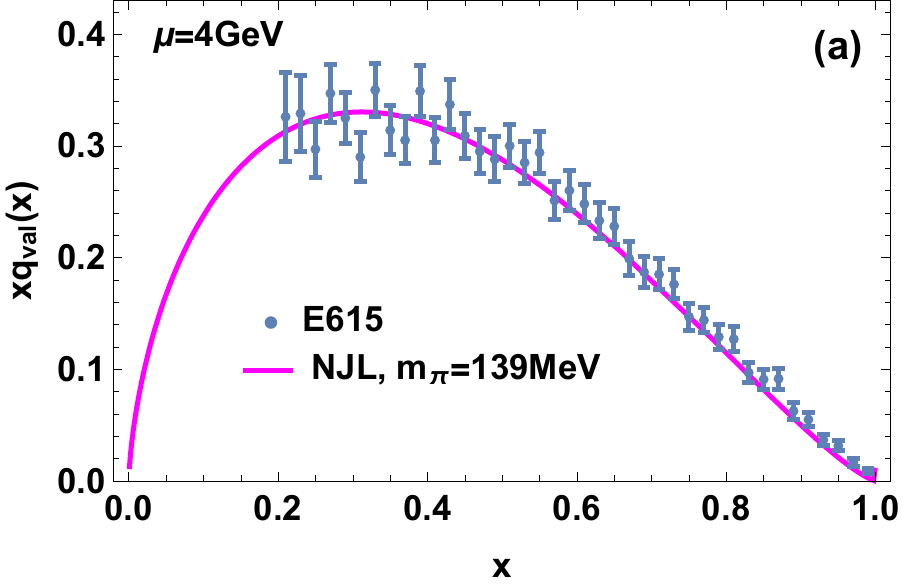} \\
\includegraphics[angle=0,width=0.35 \textwidth]{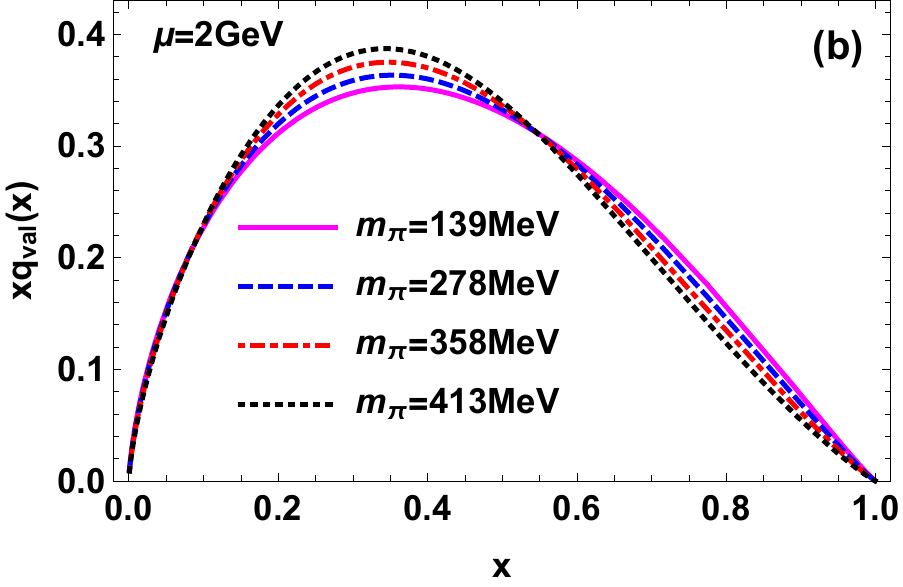}
\end{center}
\vspace{-5mm}
\caption{(a) Valence quark distribution in the NJL model evolved to the scale of 4~GeV corresponding to the E615 Fermilab data (points).
(b) LO valence quark distribution in the NJL model (multiplied with
$x$) evolved to the scale of 2~GeV corresponding to the lattice
simulations, at various values of the pion mass, including those used in \cite{Sufian:2020vzb}. \label{fig:val}}
\end{figure}

\begin{figure*}[tb]
\begin{center}
\includegraphics[angle=0,width=0.38 \textwidth]{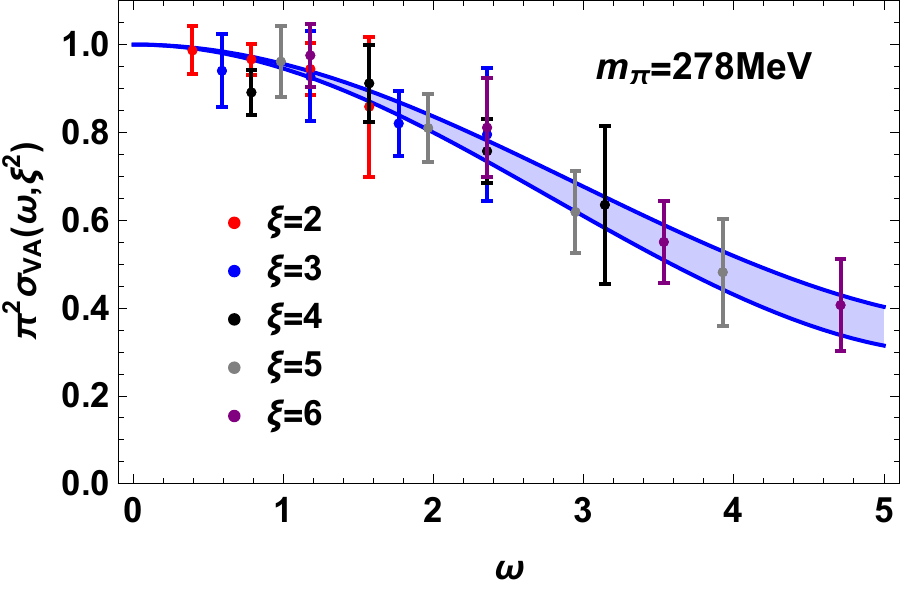} \hspace{.05\textwidth} \includegraphics[angle=0,width=0.38 \textwidth]{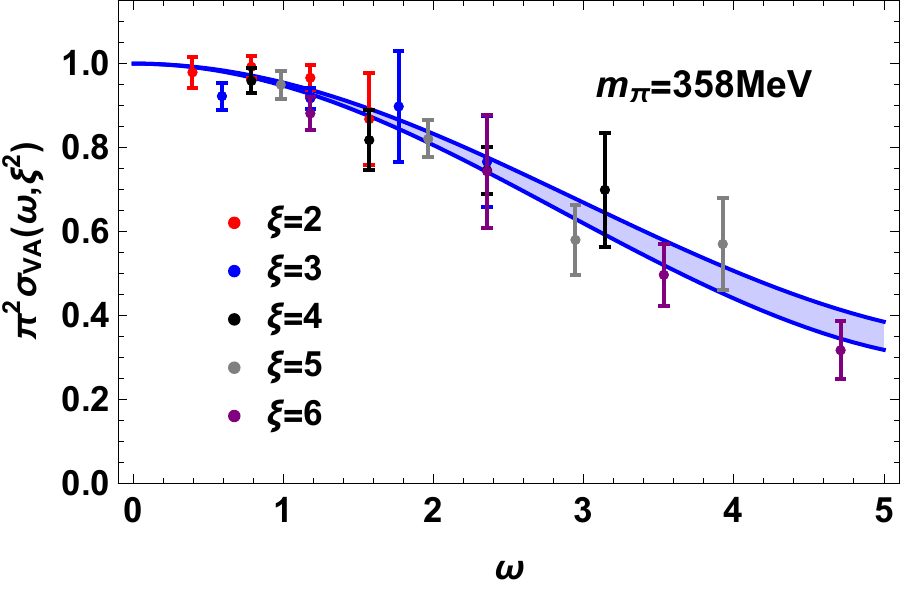}\\
\includegraphics[angle=0,width=0.38 \textwidth]{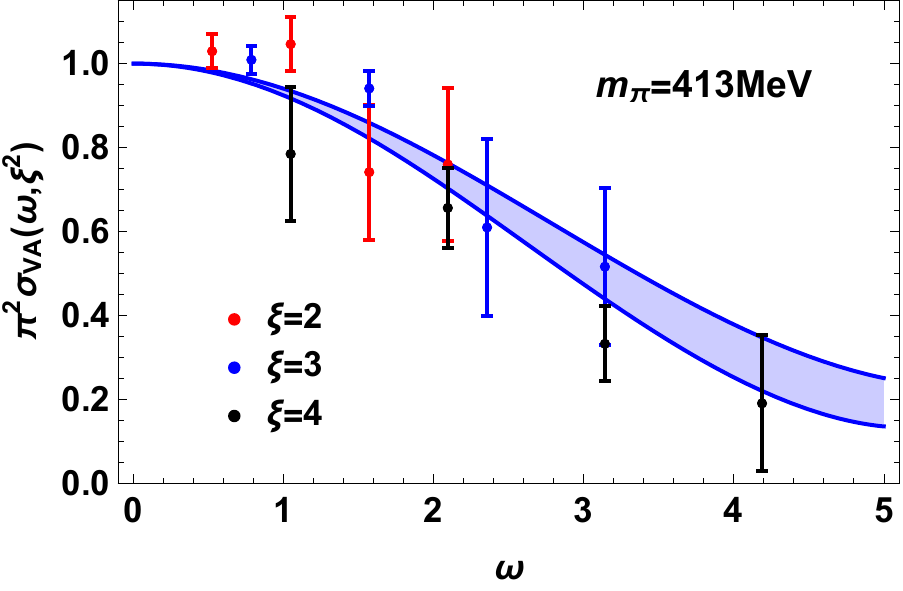} \hspace{.05\textwidth} \includegraphics[angle=0,width=0.38 \textwidth]{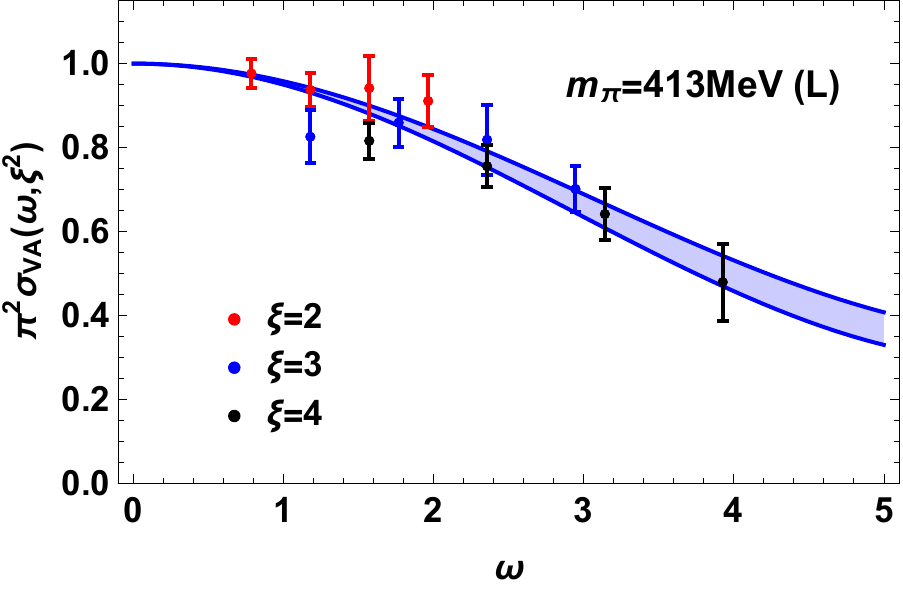}
\end{center}
\vspace{-5mm}
\caption{The good lattice cross section $\sigma_{VA}(\omega, \xi^2)$ from the NJL model evolved to the scale of the lattice data (bands),
compared to four sets of the lattice data of ~\cite{Sufian:2020vzb} (points),
plotted as functions of the Ioffe time $\omega=p \cdot \xi$. The width
of the band reflects the uncertainty in the lattice data entering the
fit of the evolution ratio $r$, which is the only adjustable parameter of the model. \label{fig:sVA}}
\end{figure*} 

In our analysis we use the pion PDF at the (yet to be determined) quark model
scale $\mu_0$, obtained from the Nambu--Jona-Lasinio (NJL) model~\cite{Davidson:1994uv,Davidson:2001cc,Weigel:1999pc}
or the Spectral Quark Model~\cite{RuizArriola:2003bs,RuizArriola:2003wi}, which (in the strict chiral limit of $m_\pi=0$) yield
\begin{eqnarray}
q_{\rm val} (x; \mu_0) = 1 \qquad {\rm for} \qquad 0 \le x \le 1, \label{eq:pdf1}
\end{eqnarray}
such that in the chiral limit we find the simple result 
\begin{eqnarray}
\sigma_{VA} ( \omega, 0 , 0 )|_{\mu_0} = \frac1{\pi^2} \frac{\sin \omega}{\omega}.
\end{eqnarray}
Chiral corrections in NJL, which are small for the physical value of $m_\pi$ and moderate for the values used on the lattice, can be 
readily evaluated from the formula
\begin{eqnarray}
q_{\rm val} (x; \mu_0) =
\frac{\left . \int d^2k_\perp \frac{k_\perp^2+M^2}{[k_\perp^2+M^2-m_\pi^2 x(1-x) ]^2} \right |_{\rm reg}}
       {\left . \int d^2k_\perp \frac{1}{k_\perp^2+M^2} \right |_{\rm reg}}, \label{eq:mp}
\end{eqnarray}
where $M$ denotes the constituent quark mass due to the dynamical
chiral symmetry breaking, and ``reg'' indicates the suitably chosen
regularization, needed to dispose of the hard momenta. We use here the
Pauli-Villars regularization as described
in~\cite{Schuren:1991sc,RuizArriola:2002wr}.

The pion PDF extracted from the experimental or lattice data largely
differs from Eq.~(\ref{eq:pdf1}), which reflects the disparity of the
quark model scale, $\mu_0$, and the scale corresponding to the data,
$\mu \sim Q$.  To provide a sensible comparison, as already advocated in
Ref.~\cite{Davidson:1994uv,Davidson:2001cc}, one crucially needs to evolve the
model results from the scale $\mu_0$ up to the scale $\mu$.  For the
problems of interest, focusing on moderate values of $x$, one can use
the DGLAP scheme, which is supposed to work best in the intermediate
$x$ region, with $x$ neither too close to $x \to 0$ nor $x \to 1$ (we
return to this point in the next section).

In chiral quark models, the valence quarks carry by definition (as the only degrees of freedom) $100\%$ of the pion's
momentum, namely
\begin{eqnarray}
\langle x \rangle_{\mu_0} = 1.
\end{eqnarray}
This condition can be used to determine numerically the {\em quark
model scale} $\mu_0$, if we know the momentum fraction carried by the
valence quarks at some other scale $\mu$, since with the DGLAP
evolution (cf. Eq.~(\ref{eq:momevo})) $\langle x \rangle_\mu = r^{\gamma_1^{(0)} /
2 \beta_0} \langle x \rangle_{\mu_0}$.  Due to the positivity of
$\gamma_1^{(0)}=\frac{64}{9}$, the evolution ratio $r$ and $\langle x \rangle_\mu $ also
decreases with $\mu$. This simply reflects the fact that some
momentum is carried by the radiatively generated gluons.  To fix
$\mu_0$, we adopt the method used in our previous works, taking that
at $\mu=2$~GeV the valence quarks carry $47\pm 2$\% of the total
momentum of the pion, as follows from~\cite{Sutton:1991ay} (see also
\cite{Gluck:1991ey,Gluck:1999xe}). At LO the
scale turns out to be $\mu_0 = 313^{+20}_{-10}$~MeV, with the
corresponding coupling $\alpha(\mu_0)/2\pi=0.34(4)$, and the evolution
ratio $r=0.15(2)$.

The result of the evolution for the PDF from the initial condition at $\mu_0$ provided with Eq.~(\ref{eq:mp})
is shown in Fig.~\ref{fig:val}(a), where we compare the model valence PDF of the pion to the experimental
extraction at LO from the E615 Fermilab data~\cite{Conway:1989fs} at
the scale $\mu= 4$~GeV.  We note that the experimental extraction of
the pion PDF was recently carried out by~\cite{Barry:2018ort}, as well
as within the xFitter framework~\cite{Novikov:2020snp}, showing
consistency at LO.  As we see from Fig.~\ref{fig:val}(a), the
agreement of the model and the data is quite remarkable. 
We note that the results of~\cite{deTeramond:2018ecg} obtained in light-front holographic QCD model are not far from ours.  The NLO
effects are below the 10\% correction level~\cite{Davidson:2001cc}.

The above discussion obviously suggests proceeding in a similar
fashion for $\sigma_{VA}$, namely, to implement the
QCD evolution on the quark model result and compare to the lattice results. This implies extrapolation to
the infinite volume limit on the lattice, but at a finite lattice spacing, embodying
the operating resolution at the corresponding wavelength $a \sim
1/Q$. In addition, one needs to adjust the pion mass in the model to
the lattice values.  The effect of such a change of $m_\pi$ for $q_{\rm val}$ 
is shown in Fig.~\ref{fig:val}(b) and, as we can see, is visible but moderate for the probed values of $m_\pi$.

\begin{figure}[tb]
\begin{center}
\includegraphics[angle=0,width=0.4 \textwidth]{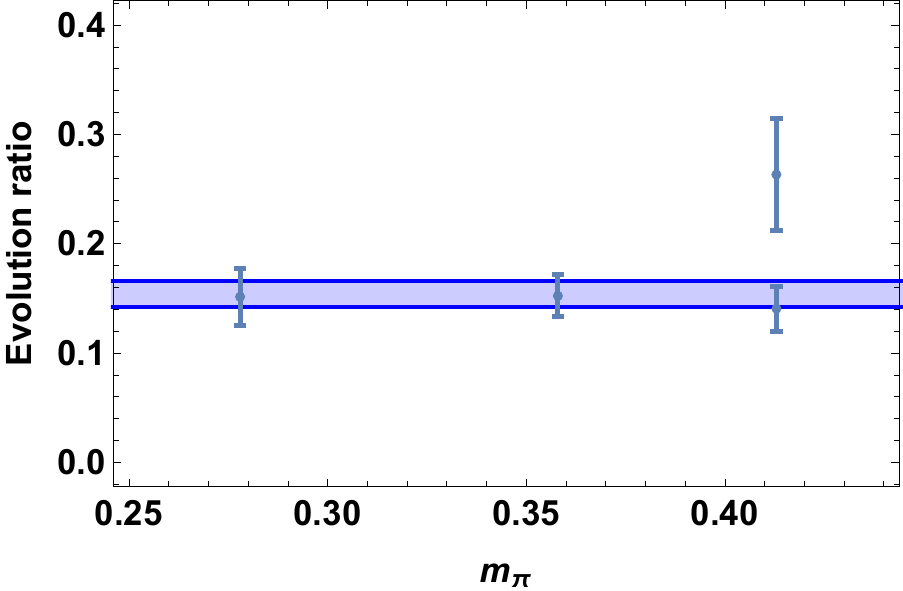} 
\end{center}
\vspace{-5mm}
\caption{Evolution ratio obtained from fitting the NJL model results to four sets of the lattice data \cite{Sufian:2020vzb} (points) shown in Fig.~\ref{fig:sVA}. 
The band indicates the weighted average with uncertainty 
reflecting the errors of the data. \label{fig:evrat}} 
\end{figure}

In Fig.~\ref{fig:sVA} we plot $\sigma_{VA}$ 
evolved to a high scale for the cases of different $m_\pi$. This is effectively performed by fitting the
evolution ratio, Eq.~\ref{eq:evol-r}, to the lattice results. As we can
see, the quality of the fits is satisfactory. The corresponding values of
$r$ are displayed on Fig.~\ref{fig:evrat}. They are consistent within the error bars, with the 
lattice data ``413L''~\cite{Sufian:2020vzb} acceptably away by 2 standard deviations from the mean.

Our weighted fit for the different pion masses allows us to determine
the evolution ratio to be $r=0.15(1)$.  Using the central value of $\mu_0=313$~MeV, determined in previous works, we infer the lattice scale
$\mu= 2.0 (3)$~GeV, which is the same as the value used in~\cite{Sufian:2020vzb}, and compatible with the lattice
spacing $a \sim 1/\mu \sim 0.1$~fm. We note that using the NLO perturbative 
kernel given in  Eq.~(10) of~\cite{Sufian:2020vzb}, one finds that for the probed values of the 
separation $\xi$ this scale is high enough such that the NLO corrections to $\sigma_{VA}$ are small.

\section{Generic analysis \label{sec:gen}}

\subsection{Moments content of lattice cross sections}

\begin{figure}[tb]
\begin{center}
\includegraphics[angle=0,width=0.4 \textwidth]{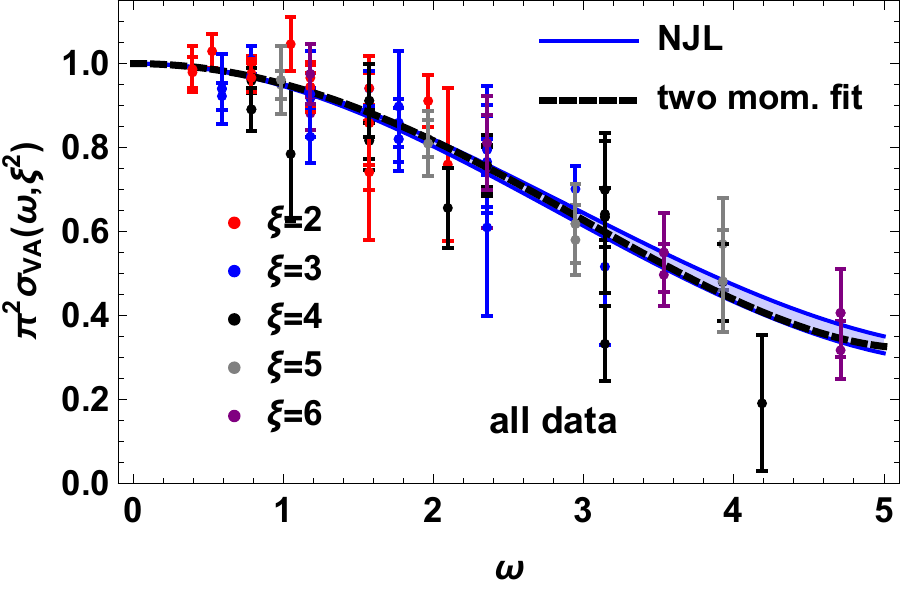} 
\end{center}
\vspace{-5mm}
\caption{Simultaneous fit of $\sigma_{VA}(\omega, \xi^2)$ to all lattice data with NJL model results 
evolved to the lattice scale of 2~GeV (band), and with a schematic model with two moments 
$\langle x^2 \rangle$ and $\langle x^4 \rangle$ treated as free parameters (dashed line). \label{fig:comb}} 
\end{figure}

An overall perspective on our study can be obtained from Fig.~\ref{fig:comb}, where
we present a combined fit of $\sigma_{VA}(\omega, \xi^2)$ to all
lattice data with NJL model results evolved to the lattice scale of 2 GeV (band). We
also show the results of a schematic model with just two moments
$\langle x^2 \rangle$ and $\langle x^4 \rangle$ treated as free
parameters:
\begin{eqnarray}
\sigma_{VA}(\omega)=1-\langle x^2 \rangle \omega^2/2! + \langle x^4 \rangle \omega^4/4!\,. \label{eq:sch}
\end{eqnarray}
As we can see, the agreement of the schematic fit with the model is remarkable.  It
also shows that for the plotted range in the Ioffe time, $0 \le \omega \le 5$, the corrections to the
schematic model from including higher moments, $\langle
x^6 \rangle, \langle x^8 \rangle$ etc., are negligible.
One may note from Table~\ref{tab:fitpar} that taking more moments as
free parameters in the schematic fit generates an overfitting effect, with
higher moments compatible with zero, and increasing errors on $\langle
x^2 \rangle$ and $\langle x^4 \rangle$. We note that the values from the schematic model~(\ref{eq:sch}) are compatible
within uncertainties with the results of~\cite{Joo:2019bzr} obtained from the Ioffe-time pseudo-PDFs.

\begin{table}
\caption{Lowest moment of the valence parton distribution of the pion at the scale 2~GeV, obtained 
from a simultaneous fit to all lattice samples from \cite{Sufian:2020vzb}. In the NJL model the only free parameter is the evolution ratio, 
which determines all $\langle x^n \rangle$ moments. In models labeled ``$m$ mom'', the $m$ lowest even moments (starting from 2) are treated as independent parameters, while 
the higher ones are set to zero. \label{tab:fitpar}} 
\vspace{3mm}
{\small
\begin{tabular}{|c|llll|} \hline
model   & $\langle x^2 \rangle$ & $\langle x^4 \rangle$ & $\langle x^6 \rangle$ & $\langle x^8 \rangle$ \\ \hline
NJL        & 0.106(5)                      &  0.036(2)                     &  0.017(1)                     & 0.010(1) \\ \hline
2 mom. & 0.099(7)                      &  0.022(5)                     & --                                 & --            \\
3 mom. & 0.101(12)                    &  0.026(25)                   & 0.005(27)                    & --            \\
4 mom. & 0.102(19)                    &  0.030(77)                   & 0.016(215)                  & 0.018(318) \\ \hline
\end{tabular}
}

\end{table}

Remarkably, the good lattice cross section method generates a stable
result for the $\langle x^2 \rangle$ moment, with an error of the
order of a few percent, and also a quite reliable estimate for the
$\langle x^4 \rangle$ moment, with accuracy of 25\%, provided higher
moments are ignored to avoid the overfitting effect.

\subsection{Aspects of the $x \to 1$ behavior}

Recent lattice
calculations~\cite{Zhang:2018rls,Chen:2019lcm,Izubuchi:2019lyk,Shugert:2020tgq}
have been undertaken with the hope to settle the long-lasting
discussion of the $x\to 1$ behavior of the PDF of the pion. The
advocated $\sim(1-x)^2$ behavior in the $x \to 1$ limit, found long
ago from the QCD counting rules~\cite{Farrar:1975yb}, is a feature not
manifest in older~\cite{Conway:1989fs} or
newest~\cite{Novikov:2020snp} extractions, and a complete reanalysis (see, e.g.,~\cite{Wijesooriya:2005ir})
and/or new experiments remain yet to be done. There are, in
particular, renormalization issues regarding elimination of scheme
ambiguities~\cite{Brodsky:1982gc} which imply the replacement
$\alpha_s(Q^2) \to \alpha_S (Q^2(1-x))$, innocuous for $x \ll 1$ but
speeding up the evolution and ultimately hitting the infrared
singularity at $x\to 1$. Another more recent scenario concerns the
inclusion of the soft-gluon resummation effects~\cite{Aicher:2010cb},
which should provide the expected counting rules behavior.  The
inclusion of these effects in order to extract the PDFs from the
Drell-Yan data should also provide a visible $(1-x)^2$ behavior in the
data (see, e.g., Ref.~\cite{Ding:2019qlr} for a discussion and
quantitative comparison).

To shed some light on this point from the lattice perspective, we have
tried several $x \to 1$ power behaviors fixing the two lowest moments
analyzed above to their fitted values and ranging between a good
description and a blatant disagreement with the E615 Fermilab
experiment~\cite{Conway:1989fs}. We find that in either case
$\sigma_{AV}$ shows no sensitivity for $\omega \le 5$, in accord with
our schematic model analysis. Thus, the insight from  the $\sigma_{VA}$ lattice
cross section into the $x \to 1$ region would require going to much higher
$\omega$ values. Similar remarks regarding the accessibility of the $x\to 1$ limit
apply also to other lattice determinations of the pion PDF.

\section{Conclusions}

In this paper we have considered a chiral quark model evaluation of
the good cross section in the Euclidean space in the vector - axial vector
channel for the pion. Our results, after the necessary 
QCD evolution (carried out at LO, with next-to leading order analysis left for future work) 
from the quark model scale to a finer lattice resolution scale $a$, display 
a comfortable agreement with the lattice data in the Ioffe-time region $\omega \le 5$, probed by
the lattice simulations at different pion masses~\cite{Sufian:2020vzb}. 
The values of the evolution ratio parameter obtained from our fits are 
perfectly compatible with our earlier estimations based on other observables, showing the 
universality of the {\em quark model + QCD evolution} scheme.
Our results also exhibit a weak dependence on the value of the
pion mass, in accordance to the lattice studies. 

We have also stressed that while the agreement in this particular model case is
predetermined by its successful reproduction of the relevant lowest moments of the
pion's PDF, the current lattice cross section data would need to be extended
well beyond the $\omega \sim 5$ region to access the $x \to 1$ kinematics, which poses a challenge.

An interesting future outlook, both on the lattice and theoretical model sides,  would involve analysis of other probing operators in lattice
cross sections, such as 
the energy-momentum tensor, which would provide complementary information on the odd moments of the PDF of the pion.

\bigskip

We cordially thank the authors of~\cite{Sufian:2020vzb} for providing
us the lattice data from the figures of their paper. 
We are grateful Micha\l{} Prasza\l{}owicz for helpful comments.

This work was supported by the Polish National Science Centre (NCN)
Grant 2018/31/B/ST2/01022 (WB), the Spanish Ministerio de Economia y
Competitividad and European FEDER funds (grant FIS2017-85053-C2-1-P)
and Junta de Andaluc\'{\i}a grant FQM-225 (ERA).

\end{document}